\documentclass[journal=jacsat,manuscript=article]{achemso}
\usepackage{graphicx}
\usepackage{epstopdf}
\usepackage{bm,amsmath,amssymb} 
\usepackage{color, soul} 
\usepackage{dcolumn}   
\usepackage{multirow}

\title{Two-dimensional multiferroic metal with  voltage-tunable magnetization and metallicity}

\author{Xu Duan}
\altaffiliation{These two authors contributed equally}
\affiliation{Fudan University, Shanghai 200433, China}
\alsoaffiliation{School of Science, Westlake University, Hangzhou, Zhejiang 310024, China}
\author{Jiawei Huang}
\altaffiliation{These two authors contributed equally}
\affiliation{Zhejiang University, Hangzhou, Zhejiang 310058, China}
\alsoaffiliation{School of Science, Westlake University, Hangzhou, Zhejiang 310024, China}
\author{Bin Xu}
\affiliation{Jiangsu Key Laboratory of Thin Films, School of Physical Science and Technology, Soochow University, Suzhou 215006, China}
\author{Shi Liu}
\email{liushi@westlake.edu.cn}
\affiliation{School of Science, Westlake University, Hangzhou, Zhejiang 310024, China}
\alsoaffiliation{Institute of Natural Sciences, Westlake Institute for Advanced Study, Hangzhou, Zhejiang 310024, China}
\alsoaffiliation{Key Laboratory for Quantum Materials of Zhejiang Province, Hangzhou Zhejiang 310024, China}

\begin{document}
\newpage
\begin{abstract}{ 
We design a multiferroic metal that combines seemingly incompatible ferromagnetism, ferroelectricity, and metallicity by hole doping a two-dimensional (2D) ferroelectric with high density of states near the Fermi level. The strong magnetoelectric effect is demonstrated in hole-doped and arsenic-doped monolayer $\alpha$-In$_2$Se$_3$ using first-principles calculations. Taking advantage of the oppositely charged surfaces created by an out-of-plane polarization, the 2D magnetization and metallicity can be electrically switched on and off in an asymmetrically doped monolayer. The substitutional arsenic defect pair exhibits an intriguing electric field-tunable charge disproportionation process accompanied with an on-off switch of local magnetic moments. The charge ordering process can be controlled by tuning the relative strength of on-site Coulomb repulsion and defect dipole-polarization coupling via strain engineering. Our design principle relying on no transition metal broadens the materials design space for 2D multiferroic metals.}
\end{abstract}

\maketitle
\newpage
Understanding the emergent phenomena arising from the dimensionality reduction has a long history in science and remains a fascinating topic. The research on two-dimensional (2D) materials, officially embarked by the seminal work of isolating graphene from bulk graphite~\cite{Novoselov04p5696}, is at the frontier of condensed matter physics and materials science, driven by the potential of these ultrathin atomic crystals for next-generation electronics with superior performances~\cite{Geim13p419,Houssa16p17}. It is now well recognized that 2D materials can adopt a broad range of electrical properties~\cite{Grazianetti20p1900439,Das15p1,Liu14p4033,Xi15p765,Cao18p43} including ferromagnetism and ferroelectricity. The famous Mermin-Wagner theorem derived rigorously from an isotropic Heisenberg model rules out the existence of long-range spin or crystalline order in 1D or 2D systems~\cite{Mermin66p1133}. Experimental and theoretical studies on quasi-2D magnetic compounds such as Rb$_2$MnF$_4$~\cite{Birgeneau70p2211}, ultra thin film magnetization~\cite{Kohlhepp92p12287}, and more recently van der Waals (vdW) 2D magnets such as CrI$_3$~\cite{Huang17p270,Gong17p265} highlighted the importance of anisotropic exchange interactions for the emergence of 2D magnetism. Meanwhile, an increasing number of 2D ferroelectrics (FEs) were discovered by experiments or by first-principles calculations based on density functional theory (DFT)~\cite{Belianinov15p3808,Zhou17p5508,Chang16p274,Yuan19p1775,Ding17p14956}.

The discovery of ferromagnetism and ferroelectricity in 2D systems naturally stimulated strong interest in the search of 2D multiferroics that combine these two ferroic orders~\cite{Lu19p653,Wei20p012005}. The design principles reported in literatures often lead to type-I multiferroic in which the magnetism and ferroelectricity have different origins~\cite{Luo17p235415,Ai19p1103}. 
For example, a 2D magnet can be functionalized covalently to induce the break of inversion symmetry~\cite{Wu13p081406}. In monolayer transition metal phosphorus chalcogenides (TMPCs)-CuMP$_2$X$_6$ (M=Cr, V; X= S, Se), the out-of-plane electric polarization is induced by the Cu off-center displacement while the magnetism stems from the indirect exchange interaction between transition metal atoms~\cite{Qi18p043102,Lai19p5163}. The main issue for type-I 2D multiferroic is the weak coupling between ferroelectric and ferromagnetic orders, leading to small magentoelectric effect. A possible solution is to construct a bilayer vdW heterostructure using ferromagnetic and ferroelectric monolayers. Recent DFT studies on Cr$_2$Ge$_2$Te$_6$/In$_2$Se$_3$ indeed demonstrated electrically-switchable magnetism~\cite{Gong19p2657}. Wu et al. proposed a series of ion-intercalated vdW bilayer (e.g., halogen-intercalated phosphorene bilayer and Co-intercalated MoSe bilayer) in which the magnetism is tunable via ion migration under an external electric field~\cite{Yang17p11506,Tu19p1800960}. It is highly desirable to obtain monolayer atomic crystals with strong magentoelectric response.

In this work, we suggest a different route to 2D multiferroic that requires no transition metal. Our starting point is the unique band dispersions presented in 2D materials: a Mexican-hat band edge (MHBE) associated with a van Hove singularity (VHS) characterized by a 1/$\sqrt{E}$ divergence in the density of states (DOS). Previous studies have already shown that adjusting the Fermi level ($E_F$) through the van Hove singularity can induce Stoner-type electronic instability and spontaneous time-reversal symmetry breaking~\cite{Cao15p236602,Seixas16p206803}. We note that the Stoner criterion does not explicitly require a MHBE; the central requirement is the presence of pronounced peak in the DOS that could well arise from other factors such as flat bands. Following that, a convenient approach to realize both magnetism and ferroelectricity in a monolayer is to charge dope 2D FEs with strong peaks in the DOS near $E_F$. An immediate concern is whether the charge doping will destroy the long-range polar ordering as the ferroelectricity and metallicity are often considered mutually exclusive. Recent studies on ``polar metal" and ``ferroelectric metal" demonstrated that the atomic distortions that break the inversion symmetry can be quite persistent against both electron and hole doping in bulk and 2D~\cite{Kolodiazhnyi10p147602, Wang12p247601,Shi13p1024,Fujioka15p13207,Puggioni15p087202,Kim16p68,Filippetti16p11211,Benedek16p4000,Cao18p1547,Fei18p336,Lu19p227601,Sharma19peaax5080}. It is thus viable to drive a 2D FE with singularities in the DOS to a magnetic state through appropriate charge doping. Moreover, for a 2D FE with out-of-plane polarization ($\mathcal{P}_{\rm OP}$), the built-in depolarization field ($\mathcal{E}_d$) running against $\mathcal{P}_{\rm OP}$ (Fig.\,1a) controls the spatial distribution of spin-polarized carriers, allowing electrical-control of magnetism. The proposed design principle does not require the presence of magnetic transition metals.
 
Here we focus on monolayer $\alpha$-In$_2$Se$_3$ which exhibits both out-of-plane and in-plane electric polarization~\cite{Ding17p14956, Zhou17p5508} due to the displacement of central Se atom relative to the top and bottom In-Se layers (Fig.\,1a). The 2D ferroelectricity in $\alpha$-In$_2$Se$_3$ was first predicted in theory using DFT, followed by experimental confirmations~\cite{Cui18p1253,Xue18p4976,Xiao18p227601, Poh18p6340,Xue18p1803738}.  The band structure computed with DFT (Fig.\,1b), using generalized gradient approximation of the Perdew-Burke-Ernzerhof (PBE) type reveals a MHBE as the the valence band maximum (VBM) is located away from the $\Gamma$ point and is 0.1~eV higher in energy, consistent with previous studies~\cite{Huang20p165157,Li20p234106}. We use monolayer $\alpha$-In$_2$Se$_3$ as a model system to study the hole-doping-induced ferromagnetism via VHS and the coupling between ferromagnetism, ferroelectricity, and metallicity in 2D.
 
We first examine the magnetic and ferroelectric properties of monolayer $\alpha$-In$_2$Se$_3$ as a function of hole carrier density ($p$) at the PBE level. It is noted that a high-density $k$-point grid (at least $60\times60\times1$) and a tight convergence threshold for self-consistency ($1.0\times 10^{-9}$~Ry) were required to fully converge the value of averaged magnetic moment per carrier $\mu_p$ (see Computational Methods) because this quantity depends sensitively on the fine details of the DOS near $E_F$~\cite{Cao15p236602}. As shown in Fig.\,1c, the ferromagnetism in monolayer $\alpha$-In$_2$Se$_3$ emerges when $p$ exceeds a critical value of $\approx13.2\times 10^{13}$/cm$^2$ (0.19 hole per 5-atom hexagonal unit cell). Meanwhile, the spin polarization energy defined as the energy difference between the nonmangetic and ferromagnetic states, $E$(NM)$-$$E$(FM), is positive, confirming a spontaneous break of time reversal symmetry upon hole doping. The 2D magnetism eventually disappears at a high carrier density ($p>18.8\times 10^{13}$/cm$^2$, 0.27 hole per unit cell). Different from previous studies on hole-doped monolayer such as GaSe, $\alpha$-SnO, and $\alpha$-In$_2$Se$_3$~\cite{Cao15p236602,Seixas16p206803,Liu21p072902}, $\mu_p$ does not reach the value of 1.0~$\mu_B$/carrier over the whole range of carrier density studied here. The spin-polarized electronic band structure and DOS at $p=14.6\times 10^{13}$/cm$^2$ (Fig.\,1d) reveals a ferromagnetic metallic state characterized by the splitting of the spin-up and spin-down bands of $\approx$42~meV for the topmost valence band. Since both spin-type bands cross $E_F$, $\mu_p <1\mu_B$ is an expected outcome. {The computed Stoner parameter $I$ and $N(E_F)$ are 0.53 and 4.27, respectively, clearly satisfying the Stoner criterion $IN(E_F) > 1$ (see details in Supplemental Material).}

The break of spatial inversion symmetry is quantified by the atomic displacement of the central Se layer along the out-of-plane direction ($d_z$) with respect to the geometric center of the monolayer. We find that the magnitude of $d_z$ decreases linearly with $p$ while maintaining a relatively large value throughout, indicating a robust polar ordering against hole doping. The $\mathcal{P}_{\rm OP}$ creates a pair of oppositely charged surfaces, denoted as $Q^+$ and $Q^-$. A direct consequence of persistent $\mathcal{P}_{\rm OP}$ and the associated $\mathcal{E}_d$ is the spatially asymmetric distribution of spin-polarized electron density ($\Delta \rho = \rho {\uparrow} - \rho {\downarrow} $): the spin-polarized hole carriers accumulate at the $Q^-$ surface. The switch of $\mathcal{P}_{\rm OP}$ will lead to the redistribution of hole carriers and an on-off switch of the surface magnetization (see inset in Fig.\,1d): in both cases only the $Q^-$ surface is magnetized. Considering that the electric field-reversible polarization is a defining feature for a ferroelectric~\cite{Lines77}, we compute the generalized Born effective charges along the out-of-plane direction, $Z^*_{33} = \partial{\mathcal{F}_3}/\partial\mathcal{E}_{3}$, where $\mathcal{F}$ is the atomic force due to a macroscopic electric field $\mathcal{E}$. Our previous studies on ferroelectric metal shows that $Z^*_{33}$ measures the degree of electric field screening due to mobile carriers~\cite{Ke21p11508}. It is found that when $\mu_p$ reaches the peak value at $p=14.6\times 10^{13}$/cm$^2$, the magnitudes of $Z^*_{33}$ are only 30\% smaller than those in the undoped semiconducting monolayer, indicating the hole carriers can not fully screen the applied external electric field. This further suggests the polarization is still reversible in the hole-doped sample albeit demanding a higher field strength.

We now discuss the relativistic effect on the hole-induced 2D magnetism. For monolayer $\alpha$-In$_2$Se$_3$, the inclusion of spin-orbit coupling (SOC) drives an indirect-to-direct band gap transition, thus removing the MHBE and the VHS in the DOS (Fig.\,1e). Our calculations confirm that the hole doping can no longer make the monolayer magnetic in the range of $p$ studied in Fig.\,1c, proofing conversely that the electronic instability identified at the PBE level is indeed of Stoner type. The MHBE of the topmost valence band including SOC can be recovered by applying a biaxial tensile strain resulting in a reemergence of 2D magnetism upon appropriate hole doping. {For example, under a biaxial strain of 2\% (Fig.\,1e), the Stoner parameter $I$ and $N(E_F)$ are 0.47 and 5.439, respectively, for monolayer $\alpha$-In$_2$Se$_3$ at $p = 10.5 \times10^{13}$~cm$^{-2}$, indicating a Stoner-type magnetic instability.} Figure 1f reports the fully relativistic band structures of strained monolayer $\alpha$-In$_2$Se$_3$ with spins oriented out-of-plane ($m$$\parallel$$z$) and in-plane ($m$$\parallel$$y$), respectively. We find that the spins favor the out-of-plane orientation with a magnetocrystalline anisotropy energy of 0.7~meV/carrier. The total magnetization changes from 1$\mu_B$/carrier for $m$$\parallel$$z$ to 0.21$\mu_B$/carrier for $m$$\parallel$$y$. The states near $E_F$ and the Fermi surface topology also experience notable modulations upon the rotations of spin directions, referred to as ``spin-textured" band structure effect~\cite{Liao20p30}, which can be utilized to control the transport and optical properties by rotating the magnetization. 

From the study of ideally hole-doped monolayer $\alpha$-In$_2$Se$_3$, we make the following arguments. The presence of high DOS near $E_F$ is critical to drive a 2D FE to a magnetic state via charge doping at a reasonable carrier density while maintaining the polar ordering and switchability. Though the inclusion of relativistic effect may alter the band dispersions, the strain engineering can recover the needed high DOS to satisfy the Stoner criterion. The hole-doped monolayer $\alpha$-In$_2$Se$_3$ remains to be type-I multiferroic in nature: despite the electric field-tunable spatial distribution of spin-polarized charge density, the {\em total magnetization} is independent of the polarization direction. In what follows, we offer a proof-of-principle demonstration that introducing $p$-type dopants represented by arsenic (As) to one surface of the monolayer, referred to as ``asymmetric doping", can result in a diverse types of magnetoelectric responses. Given the substantial computational costs associated with fully relativistic calculations using high-density $k$-point grid, we will study the magnetoelectric effect in asymmetrically As-doped monolayer at the PBE level without SOC. 

{Introducing realistic anion-site acceptors such as As to only one surface of the monolayer brings in multiple intricately coupled factors affecting the 2D magnetism. Previous studies on $p$-type conductivity in oxide and nitride semiconductors have shown that acceptor dopants can be categorized into two types: the shallow, effective-mass acceptor and the ``atomic-like" deep acceptor with the former characterized by delocalized hole wave function and the latter featuring highlighly localized hole wave function around the dopants~\cite{Lyons14p012014}. The nature of the electronic structure of the anion-site acceptor depends on the relative energy of the impurity level with respect to the valence and conduction bands of the host material.  Since the surface bound charge state ($Q^-$ and $Q^-$) serves as the electrostatic environment of dopants, the switch of $\mathcal{P}_{\rm OP}$ could change the energies of defect bands, thus affecting the doping type. In monolayer $\alpha$-In$_2$Se$_3$, the built-in depolarization field will push the hole carriers toward the $Q^-$ surface.  Depending on whether the $Q^-$ surface is doped or not, the nature of the magnetism (itinerant vs. localized) is therefore tunable via polarization switching.}

{
We first study the electronic structure of asymmetrically doped monolayer by replacing one surface Se atom with one As atom in a $2\times2$ hexagonal supercell, corresponding to a nominal carrier concentration of $p=17.2\times 10^{13}$/cm$^2$ (assuming one As atom introduces one hole carrier).  This is actually a rather high doping concentration. Interestingly, we observe a 2D magnetism directly coupled with the direction of $\mathcal{P}_{\rm OP}$. When the polarization direction has As dopants embedded in the $Q^-$ surface (As@$Q^-$,Fig.\,2a left), the computed PBE band structure reveals a semiconducting state with a As-centered local magnetic moment (see additional details in Supplemental Material). The atomic orbital-resolved DOS reveals that the spin-down impurity level arising from As dopants lies slightly above the Fermi level. It is concluded that the As dopants in the $Q^-$ surface are better regarded as ``atomic-like" deep acceptors with localized holes. The polarization reversal changes the electrostatic environment of As dopants from $Q^-$ to $Q^+$. The resulting configuration, denoted as As@$Q^+$, is surprisingly nonmagnetic (Fig.\,2a right).}

{ Our model study (Fig.\,1c) suggests that the 2D magnetism eventually disappears at a high carrier density. Accounting for the relatively high As doping concentration studied in Fig.\,2, the absence of magnetism in the As@$Q^+$ configuration is likely due to a too high hole concentration. We investigate the electronic structure of this As-doped monolayer with added electrons. Expectedly, the reduction of the effective hole concentration ($p_{\rm eff}$) to  $p=10.3\times 10^{13}$/cm$^2$ recovers magnetic $Q^-$ surface as confirmed by the spin-polarized charge density and spin-resolved density of states (Fig.\,2b).  We then examine the magnetic properties of monolayer As-doped  $\alpha$-In$_2$Se$_3$ with $p_{\rm eff}$ tuned by doping electrons (Fig.\,2c). 
The ferromagnetism in the As@$Q^+$ configuration emergences when $p_{\rm eff}$ is roughly in the range of $5-16\times 10^{13}$/cm$^2$. By the meanwhile, the As@$Q^-$ configuration always adopts localized magnetism. The observed strong coupling between the total magnetization and polarization direction hints at a possible electrical on/off switch of 2D magnetization as well as metallicity through asymmetric As-doping at an appropriate concentration (shaded region in Fig.\,2c). Such feature can be utilized to achieve the ``magnetic reading + electrical writing" in mutiferroic-based memory~\cite{Lu19p653}.  We note that the polarization switching remains energetically feasible even at a high As doping concentration (see details in Supplemental Material). 
}

{Guided by the results shown in Fig.\,2c, we explore the electronic structure of As-doped monolayer at a lower doping concentration.} The supercell consists of $3\times3$ hexagonal unit cells in which two As atoms substitute two surface Se atoms ($p_{\rm eff} =15.3\times 10^{13}$), and the neighboring As-As distance is $\approx$7.1~\AA. {This model also allows us to study the competition between ferromagnetic (FM) and antiferromagnetic (AFM) ordering of localized spins.} When the out-of-plane polarization generates a As@$Q^-$ configuration, the monolayer is a semiconductor and each As atom acquires a local magnetic moment of 1.0$\mu_B$ (Fig.\,3a).  {It is found that the FM and AFM configurations are close in energy, with the AFM phase lower in energy by 0.8 meV/dopant, indicating weak exchange interactions (0.13 meV/pair) between local spin moments. Given the small energy difference, the  As@$Q^-$ configuration is likely paramagnetic at elevated temperatures. }

Consistent with results shown in Fig.\,2c, the switch of $\mathcal{P}_{\rm OP}$ causes a semiconductor-to-metal transition, accompanied with a transition from a localized magnetism to an itinerant magnetism (Fig.\,3c). {In the As@$Q^+$ configuration, the As-free $Q^-$ surface is  magnetized (0.5$\mu_B$/carrier)}. The exchange splitting between the topmost spin-up and spin-down valence bands is $\approx$65~meV along $\Gamma-\rm{K}$ direction (see band structures in Supplemental Material) that may support a finite-temperature ferromagnetism~\cite{Cao15p236602}. The change in the nature of magnetism not only allows an on-off switch of global magnetization at finite temperature through polarization reversal (Fig.\,3), but also strongly affects the spin-related transport properties. Our results demonstrate that As-doped $\alpha$-In$_2$Se$_3$ can serve as a versatile platform for coupled metallicity, ferroelectricity, and ferromagnetism, all controllable via voltage. 

{
It is well known that semi-local density functionals such as PBE underestimate the band gap because of the the remnant self-interaction error (SIE) originating from the Hartree term.  
By adding a portion of the exact Hartree-Fock exchange that cancels the SIE, hybrid functionals reduce electron overdelocalization and therefore improve the band gap prediction. However, popular hybrid functionals such as HSE06 often assume fixed dielectric screening, making their applications to low-dimensional materials questionable as the rapid variation of screened Coulomb interactions is not appropriated captured~\cite{Jain11p216806}. It is thus important to check whether the semiconductor-to-metal transition as well as the localized-itinerant magnetic transition in As-doped monolayer identified with PBE is robust. To address this concern, we employ newly developed pseudohybrid Hubbard density functional, Agapito-Cuetarolo-Buongiorno-Nardelli (ACBN0)~\cite{Agapito15p011006}, and the extended version (eACBN0)~\cite{Lee20p043410,Tancogne-Dejean20p155117}. The ACBN0 functional is a DFT+$U$ method with the on-site Coulomb repulsion $U$ value determined self-consistently. The eACBN0 is an extended DFT+$U$+$V$ method with $V$ representing inter-site Coulomb repulsion between the neighbouring Hubbard sites. Because both $U$ and $V$ are performed on each atom self-consistently, eACBN0 caputres the local variations of screening of Coulomb interactions. Previous benchmark studies have shown that ACBN0 and eACBN0 have better descriptions of the electronic structures of low-dimensional materials than HSE06 and GW~\cite{Lee20p043410, Huang20p165157}. Figure~\ref{compareBands} compares the PBE, ACBN0, and eACBN0 band structures of As-doped monolayer $\alpha$-In$_2$Se$_3$ at the same doping concentration of $15.3\times 10^{13}$/cm$^2$. For the  As@$Q^-$ configuration, all three DFT methods predict a semiconducting state with localized spins, with eACBN0 giving the largest band gap followed by ACBN0 and PBE. Consistently, the As@$Q^+$ configuration appears to be a half metal, though eACBN0 predicts a larger splitting of the spin-up and spin-down bands. Moreover, the formation energy of As$_{\rm Se}$ is 1.85~eV, comparable with formation energies of typical dopants in bulk semiconductors and 2D materials (see details in Supplemental Material). These results confirm the validity of our proposed design principle based on asymmetric As doping. 
}

Finally, we report an intriguing electric field-tunable charge disproportionation process between two neighboring As dopants at the $Q^-$ surface assisted by the in-plane formal polarization of $\alpha$-In$_2$Se$_3$. Assuming carriers are localized around the dopants, the charge neural defect $\rm As_{Se}^\times$ (in Kr\"oger-Vink notation) contains one hole.  For two $\rm As_{Se}^\times$ defects in close proximity to each other, following charge disproportionation process is possible:
\[
\rm 2As_{Se}^\times \rightarrow As_{Se}' +  As_{Se}^{\bullet}
\]
where $\rm As_{Se}'$ with a single negative charge has no hole while $\rm As_{Se}^{\bullet}$ with a single positive charge contains two holes. The two charged defects then form a defect dipole $\mu_D$, $\rm \left(As_{Se}' -As_{Se}^{\bullet} \right)^\times$. The disproportionation process to form a pair of charged defects is generally unfavored because of the on-site Coulomb repulsion term between the two holes at $\rm As_{Se}^{\bullet}$. However, when embedded in a ferroelectric medium, the defect dipole may couple strongly with the polarization through the dipolar electric field~\cite{Ren04p91}. Previous studies on dipolar impurities in bulk ferroelectrics have shown that the defect dipole prefers to align with the bulk polarization~\cite{Zhao19pe00092, Liu17p082903}. Therefore, when the energy gain from the defect dipole-polarization coupling becomes dominant over the energy penalty arising from the on-site Coulomb repulsion, the disproportionation process to form a pair of charged defects is energetically feasible. Figure~\ref{AsCD}a shows three typical configurations. Specifically, a defect dipole $\rm \left(As_{Se}^{\bullet}-As_{Se}' \right)^\times$ with collinear $\mu_D$ and in-plane polarization ($\mathcal{P}_{\rm IP}$) is denoted as $[2^+\leftarrow 0^-]$ in which the number specifies the number of localized holes, the superscript is the charge state, and the arrow represents the direction of $\mathcal{P}_{\rm IP}$. The two-fold rotational symmetry within the plane ensures $[2^+\leftarrow 0^-]$  being equivalent to  $[0^-\rightarrow 2^+]$. In comparison, $[0^-\leftarrow 2^+]$ has $\mu_D$ (pointing from $-$ to $+$) running against the in-plane polarization ($\leftarrow$), and thus has a higher energy than $[2^+\leftarrow 0^-]$. The charge neutral defect pair is denoted as $[1, 1]$ and its energy is insensitive to the direction of $\mathcal{P}_{\rm IP}$. This is the case of localized magnetism shown in Fig~\ref{twoAs}a. It is evident from Fig.\,\ref{AsCD}a that the three configurations have distinct magnetic patterns formed by local spin moments. 

We are able to obtain the optimized structures of $[2^+\leftarrow 0^-]$ and $[1,1]$ states in asymmetrically doped monolayer. The major structural differences between the two states are the size of the In$_3$ triangle around As: the positive (negative) $\rm As_{Se}^{\bullet}$ ($\rm As_{Se}'$) repel (attract) neighboring In atoms, while the size of the In$_3$ triangle around $\rm As_{Se}^\times$ is inbetween (Fig.\,4\ref{AsCD}a). This is similar to the varying volumes of oxygen octahedral around Fe$^{3+}$ and Fe$^{5+}$ in charge-ordered La$_{1/3}$Sr$_{2/3}$FeO$_3$~\cite{Park19p23972}. The high-energy $[2^+\rightarrow 0^-]$ state is not stable, and is obtained by constraining the sizes of In$_3$ triangles around As dopants while relaxing all the other atoms. From the density of states (Fig.\,\ref{AsCD}b), we find that the $[1,1]$ configuration is a semiconductor while the $[2^+\leftarrow 0^-]$  is a half-metal.  Figure~\ref{AsCD}c presents the evolutions of energies and local magnetic moments at doping sites along the interpolated pathways connecting three charge states. The $[1,1]$ state has the lowest energy at $\eta =0\%$. Notably,  the relative energy between $[1, 1]$ and $[2^+\leftarrow 0^-]$ can be tuned by applying a biaxial strain. In the presence of a 1\% compressive strain, $[2^+\leftarrow 0^-]$ becomes the most favored state, enabling an on-off switch of local magnetic moments. Illustrated in Fig.\,\ref{AsCD}d, the initial state $[2^+\leftarrow 0^-]$ has site-$A$ spin polarized and site-$B$ being nonmagnetic. The reverse of the in-plane polarization changes the state to the high-energy configuration $[2^+\rightarrow 0^-]$ (equivalent to $[0^-\leftarrow 2^+]$ in Fig.\,\ref{AsCD}c) as $\mu_D$ and $\mathcal{P}_{\rm IP}$ are now antiparallel. Thermodynamically, the system will eventually relax to the state $[0^-\rightarrow 2^+]$ in which site-$B$ becomes spin polarized, thus turning off the magnetization at site-$A$ and turning on the magnetization at site-$B$.

In summary, we explored the potential of combining ferroelectricity, ferromagnetism, and metallicty in two dimensions by hole doping ferroelectrics with high density of states near the Fermi level based on first-principles density functional theory calculations. The ferroelectricity and metallicity are often considered mutually exclusive; the coupling between ferroelectricity and ferromagnetism that arise from different origins leads to a weak magnetoelectric response. Using hole and As-doped monolayer $\alpha$-In$_2$Se$_3$ as examples, we demonstrate that the emergence of 2D magnetization and its nature can be fully controlled by tuning the direction of out-of-plane polarization, leading to a genuine multiferroic metal with a strong magnetoelectric coupling. By engineering the As dopant concentration, a variety of electronic and magnetic phase transitions can be controlled electrically in 2D, an attractive feature for ultrathin and multistate memory. Finally, the in-plane polarization is found to mediate a charge disproportionation process within a defect pair. The ability to precisely control the on/off state of the local magnetic moments via voltage offers a new route to high-density multiferroic memory.  Our design principle relying on no-transition metals can be generalized to other 2D systems.  \\

{\bf{Computational Methods}}

DFT calculations are performed using \texttt{QUANTUM ESPRESSO}~\cite{Giannozzi09p395502, Giannozzi17p465901}. The exchange-correlation functional is treated within the generalized gradient approximation of the
Perdew-Burke-Ernzerhof (PBE) type~\cite{Perdew96p3865}. The GBRV ultrasoft pseudopotentials~\cite{Garrity14p446}  are used for structural optimizations and electronic structure calculations without spin-orbit coupling, and fully relativistic projector argument wave (PAW) pseudopotentials taken from Pslibrary (version 1.0.0)~\cite{DalCorso14p337} are used for electronic property calculations considering spin-orbit coupling. To model the monolayer, the simulation box with periodic boundary conditions contains a unit cell and a vacuum region of more than 15 \AA~along the out-of-plane direction ($z$ axis). Our benchmark calculations show that the structural and electronic properties of monolayer $\alpha$-In$_2$Se$_3$ are not sensitive to the dipole correction. The plane-wave energy and charge density cutoffs are set to 50 Ry and 250 Ry, respectively. We tested $k$-point grids of increasing $k$-point number from $20\times20\times1$ to $80\times80\times1$. The structural parameters reach convergence quickly with respect to the $k$-point grid density. However, for a given structure, a high-density $k$-point grid and an extremely tight convergence threshold for electronic self-consistency are needed to obtain a converged value of magnetization. It is also critical to use the tetrahedron method for Brillouin-Zone integrations because the Stoner-type magnetic instability depends sensitively on density of states near the Fermi level. Based on our benchmark calculations, for hole-doped monolayer, we use a 20$\times$20$\times$1 Monkhorst-Pack $k$-point grid for structural optimizations, and a $60\times60\times1$ $k$-point grid and a $10^{-9}$~Ry convergence threshold for electronic self-consistent calculations. The optimized tetrahedron method~\cite{Kawamura14p094515} is adopted to perform integrations over the $k$ points.
For As-doped monolayer, a 8$\times$8$\times$1 $k$-point grid is used for ionic relaxations with lattice constants fixed to the values of undoped monolayer. The electronic properties are computed with a 15$\times$15$\times$1 $k$-point grid that is enough to converge the magnetization. The electronic self-consistent convergence threshold is also set to $10^{-9}$~Ry.\\

{\bf{Acknowledgments}} X.D., J.H., and S.L. acknowledge the supports from Westlake Education Foundation, Westlake Multidisciplinary Research Initiative Center, and National Natural Science Foundation of China (52002335). The computational resource is provided by Westlake HPC Center.  

\bibliography{SL}

\newpage
\clearpage
\begin{figure}[t]
\centering
\includegraphics[scale=1]{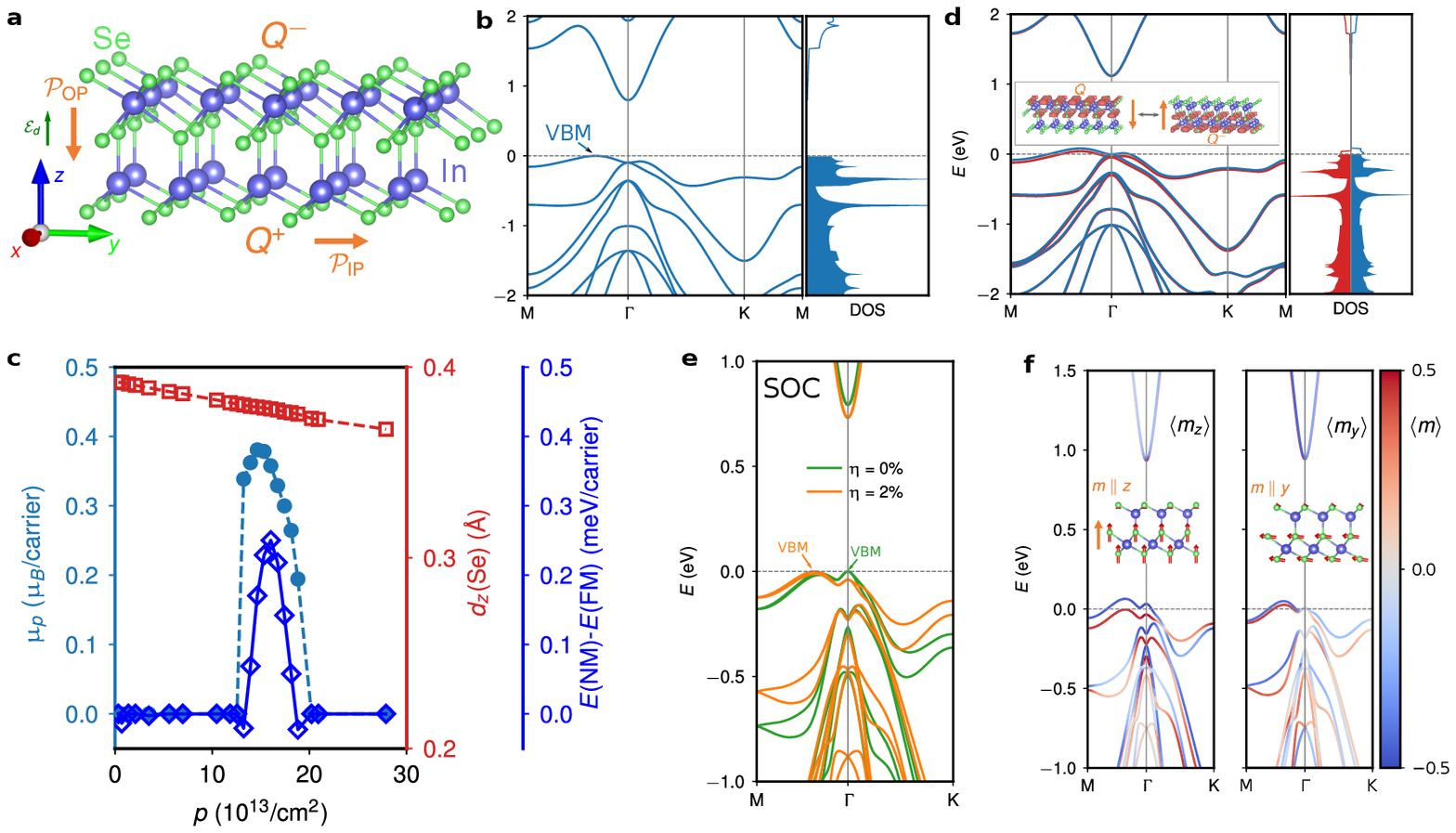}
 \caption{Electronic structures of undoped and hole-doped monolayer $\alpha$-In$_2$Se$_3$. ({\bf a}) Crystal structure of monolayer $\alpha$-In$_2$Se$_3$ with both in-plane and out-of-plane polarization. The out-plane-polarization ($\mathcal{P}_{\rm OP}$) creates oppositely charged $Q^+$ and $Q^-$ surfaces.  The depolarization field ($\mathcal{E}_{d}$) is against $\mathcal{P}_{\rm OP}$. ({\bf b}) PBE band structure and density of states of undoped monolayer. The valence band maximum (VBM) is located away from the $\Gamma$ point.  ({\bf c}) Evolution of magnetic moment ($\mu_p$ in Bohr magneton $\mu_B$) and spin polarization energy [$E$(NM)-$E$(FM)] per carrier and atomic displacement of the central Se layer ($d_z$) as a function of hole carrier density ($p$). ({\bf d}) Spin-polarized band structure and density of states at $p=14.6\times 10^{13}$/cm$^2$. The inset shows the isosurface of spin-polarized electron density.  ({\bf e}) Electronic band structures with spin-orbit coupling (SOC) of unstrained ($\eta=0$\%) and biaxially strained ($\eta=2$\%) monolayers. ({\bf f}) Fully relativistic band structures of strained ($\eta=2$\%) and hole-doped monolayer $\alpha$-In$_2$Se$_3$ with spin directions aligned out-of-plane ($m$$\parallel$$z$, left) and in-plane ($m$$\parallel$$y$, right) at $p=10.5\times 10^{13}$/cm$^2$. The color scales with the expectation value of the spin operators ($\left< m_z\right>$ and $\left< m_y\right>$) on the spinor wavefunctions.}
  \label{design}
 \end{figure}

\newpage
\begin{figure}[t]
\centering
\includegraphics[scale=1.5]{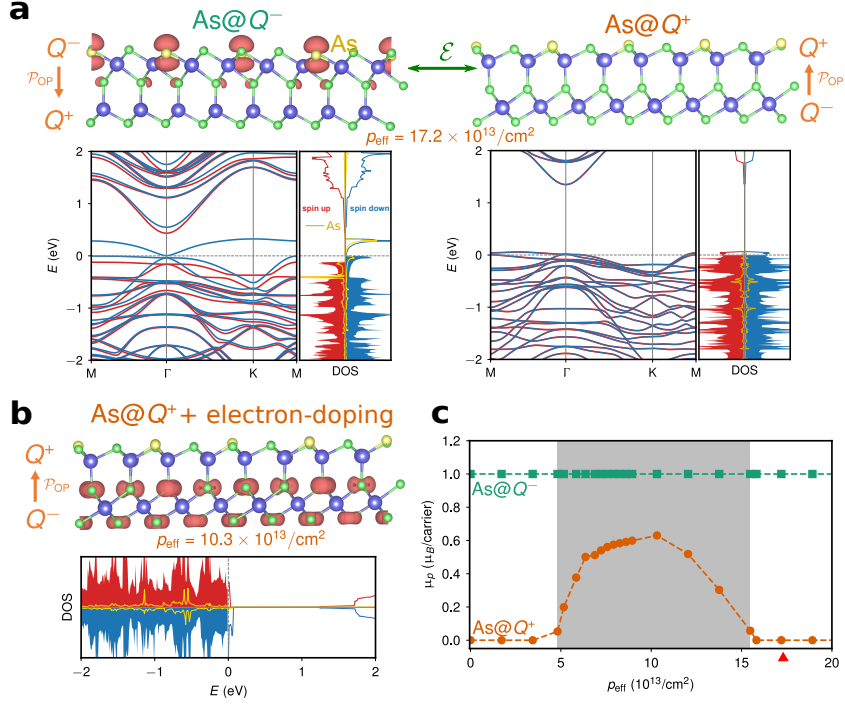}
 \caption{{Electrical on/off switch of 2D magnetization in As-doped monolayer $\alpha$-In$_2$Se$_3$. ({\bf a}) Spin-polarized electron density, band structures, and density of states corresponding to two polarization states, As@$Q^-$ and As@$Q^+$. The $2\times2$ supercell has one As atom replacing one surface Se atom, corresponding to an effective hole concentration ($p_{\rm eff}$) of $17.2\times 10^{13}$/cm$^2$. The As@$Q^-$ configuration (left) is a semiconductor with localized magnetic moments around As dopants. The As@$Q^+$ configuration (right) resulting from polarization reversal is nonmagnetic. ({\bf b}) Emergence of 2D magnetism in As@$Q^+$ configuration due to electron doping. Reducing $p_{\rm eff}$ to $10.3\times 10^{13}$/cm$^2$ by doping electrons makes the dopant-free $Q^-$ surface magnetic. The value of the isosurface is set at $6.0\times 10^{-4}$$e$/bohr$^3$. ({\bf c}) Evolution of magnetic moment per carrier ($\mu_p$) as a function of $p_{\rm eff}$ in As-doped monolayer for As@$Q^-$ and As@$Q^+$ configurations. The value of $p_{\rm eff}$ is varied by doping electrons. The red filled triangle marks the value of $p_{\rm eff}$ without electron doping. The shaded region represents possible doping concentration that enables electrical on/off switch of 2D magnetization and metallicity. 
 }}
  \label{oneAs}
 \end{figure}

\newpage 
\clearpage
\begin{figure}[t]
\centering
\includegraphics[scale=1]{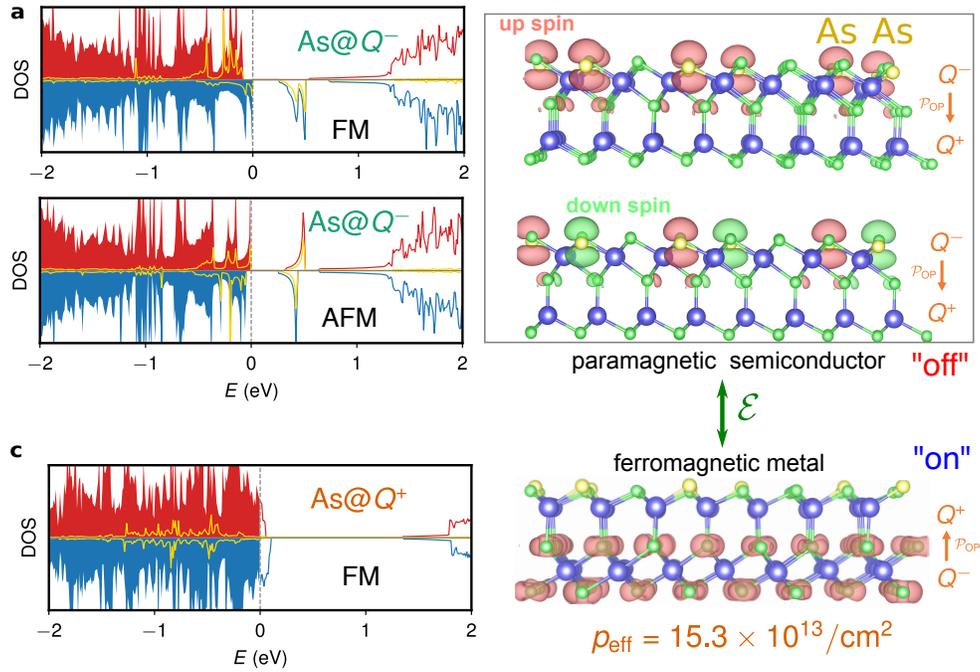}
 \caption{{Transition between localized magnetism and itinerant magnetism in As-doped monolayer $\alpha$-In$_2$Se$_3$ at a doping concentration of $15.3\times 10^{13}$/cm$^2$. Spin-resolved density of states and isosurface of spin-polarized electron density of ({\bf a}) ferromagnetic (FM) and ({\bf b}) antiferromagnetic (AFM) phases when As dopants are at the $Q^-$ surface (As@$Q^-$), and ({\bf c}) itinerant magnetism in the As@$Q^+$ configuration. It is feasible to use an electrical field to realize on/off switch of global 2D magnetization and metallicty. }}
  \label{twoAs}
 \end{figure}

\newpage 
\begin{figure*}[t]
\centering
\includegraphics[scale=1.5]{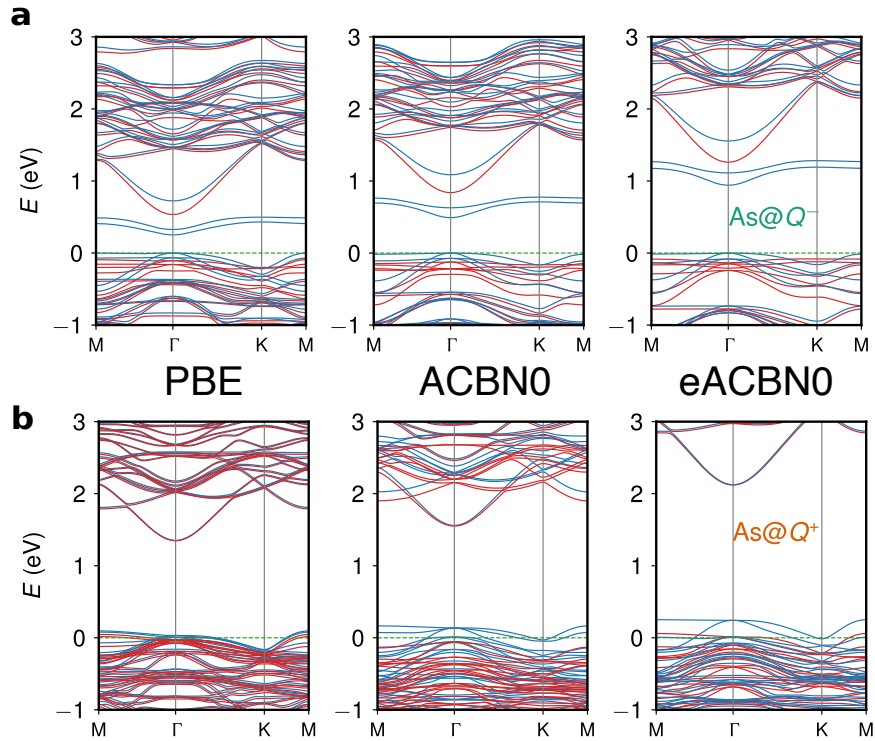}
 \caption{Band structures of As-doped monolayer $\alpha$-In$_2$Se$_3$ at a doping concentration of $15.3\times 10^{13}$/cm$^2$ for ({\bf a})  As@$Q^-$ configuration and ({\bf b})  As@$Q^+$ configuration obtained with PBE, ACBN0, and eACBN0. All DFT methods predict a semiconducting As@$Q^-$ configuration with localized spins and a half-metallic As@$Q^+$ configuration with itinerant magnetism.}
  \label{compareBands}
 \end{figure*}
 
\newpage 
\begin{figure*}[t]
\centering
\includegraphics[scale=0.9]{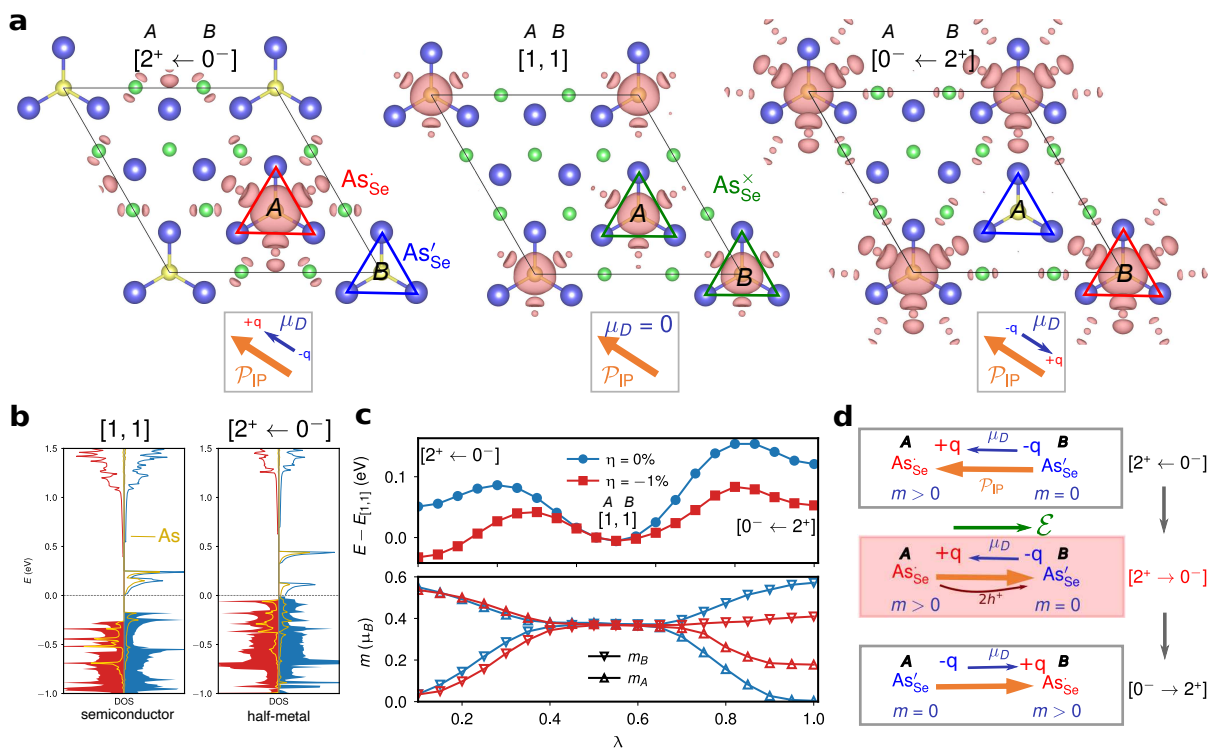}
 \caption{Charge disproportionation process between two neighboring As dopants at $Q^-$ surface mediated by the in-plane polarization ($\mathcal{P}_{\rm IP}$). ({\bf a}) Isosurfaces of spin-polarized electron density for three different charge states of As defect pairs. The size of the triangle formed by the three In atoms around As depends on the charge state of the defect. The triangle around $\rm As_{Se}^{\bullet}$ (red) is larger than that containing $\rm As_{Se}^\times$ (green), followed by the one around $\rm As_{Se}'$ (blue). ({\bf b}) Density of states for [1,1] and $[2^+\leftarrow 0^-]$. ({\bf c}) Interpolated pathway connecting three charge states under two biaxial strain conditions, $\eta=0\%$ and $\eta=-1\%$. The top panel shows the change in energy relative to $E_{[1,1]}$; the bottom panel shows the evolutions of spin moments localized at site-$A$ and site-$B$, respectively.  ({\bf d}) Schematics of electrical on/off switch of local magnetic moments by switching the in-plane polarization via an in-plane electric field $\mathcal{E}$.}
  \label{AsCD}
 \end{figure*}




\end{document}